\documentstyle[12pt]{article}  
\begin{document}

\vspace*{-10mm}  
  
\hfill IEEC/CSM-00-31  
  
\hfill quant-ph/0007035  
  
\hfill October 2000 (revised)
 
\thispagestyle{empty}

\vspace*{4mm}

\begin{center}  
  
{\LARGE \bf Quantum deletion is possible via a partial randomization 
procedure}

\vspace{4mm}  
  
\medskip

{\sc E. Elizalde}\footnote{Presently on leave at Department of Mathematics,
Massachusetts Institute of Technology, 77 Massachusetts Ave, Cambridge,
MA 02139. E-mail:  elizalde@math.mit.edu \ 
elizalde@ieec.fcr.es \ \  
http://www.ieec.fcr.es/cosmo-www/eli.html} \\  
Instituto de Ciencias del Espacio (CSIC) \\ \&
Institut d'Estudis Espacials de Catalunya (IEEC/CSIC), \\  
Edifici Nexus, Gran Capit\`a 2-4, 08034 Barcelona, Spain\\ and \\  
Departament ECM i IFAE, Facultat de F\'{\i}sica, \\  
Universitat de Barcelona, Diagonal 647,  
08028 Barcelona, Spain \\

\vspace{6mm}  
  
{\bf Abstract}  
  
\end{center}  
 
An alternative kind of deleting/erasing operation is introduced which 
differs from the commonly used {\it controlled-not}  (C-not)
conditional logical operation $-$to flip to a standard, `zero' value 
the (classical or quantum) state of 
the last copy in a chain, in a deletion process. It is completely reversible,
in the classical case, possessing a most natural cloning operation
counterpart. We call this deleting
procedure R-deletion since, in a way, it
can be viewed as a `randomization' of the standard C-not operator. It has the
remarkable property of by-passing in a simple manner the `impossibility
of deletion of a quantum state' principle,
put forward by Pati and Braunstein recently \cite{pbn1}.

\vspace{3mm}  
  

\newpage

The possibility of constructing a consistent quantum information theory 
and, what is more, 
of making practical use of its impressive potentialities (e.g., quantum 
cryptography \cite{bb1}, quantum teleportation \cite{cb1,db1}, 
quantum computing \cite{z1},  etc.) 
has attracted considerable attention in the Physics community during 
the last twenty years \cite{z1}. Almost as old is, however, the quantum 
`no-cloning theorem', due to Wooters and Zurek \cite{wz1}, and Dieks 
\cite{d1}, which prevents the replication of unknown quantum states. In spite
 of this result $-$which does not allow, in particular,  quantum information
to be amplified accurately$-$ and as it usually happens with no-go theorems,
different alternatives to circumvent this strict prohibition have appeared 
in the literature
\cite{bh1}-\cite{p1}. They reestablish, in several different ways, a sort of
consistency between the quantum theory and its possible real 
application as an information theory, which will obviously have the processes 
of copying,
storing, and retrieving of information as some of its most basic tasks.  
However, such consistency has been regained, at the very best, only
at the level  of arbitrarily good approximations in specific circumstances, 
while the  no-cloning theorem still remains as a  cornerstone in this field.
 
Recently,  Pati and  Braunstein have formulated what looks  a 
complement to this theorem, namely a proposition that can be termed as  a
quantum `no-deleting theorem' \cite{pbn1}. In the same way as the no-cloning 
theorem told us that, contrary to what happens in classical information
issues, the cloning of  quantum information cannot be taken for granted,
it also would now happen that deleting quantum information is in no way
 a trivial
operation. One might argue, at first sight, that this result sounds quite old 
and well known (see Szilard \cite{sz1} and Landauer \cite{la1}). In fact,
replacing a sequence of 0's and 1's by a perfectly ordered state of 0's
is thermodynamically costly, at the classical level. But the concept of 
deletion used in \cite{pbn1} has
 a more subtle formulation, which prevents the result from being
 trivial, and makes it much resemble the inverse of
the usual cloning operation in \cite{wz1,d1}, in the sense that it consists 
in the deletion of just one copy of a bit that is kept in several 
(at least two)
copies. In other words, this sort of deletion can only start when one has two
arbitrary, unknown but identical, bits (which can be termed `original'
and `copy').

To be precise, and for the benefit of the reader, let us here recall 
the controlled not (or C-not) operation of classical information theory,
which allows the deletion and copying of an arbitrary sequence of classical 
bits (see, for instance, Zurek \cite{zn1}). Confronted with a pair of states,
$|s_1>|s_2>$, C-not yields:
\begin{eqnarray}
&&|0>|0>\longrightarrow  |0>|0>, \ \ |0>|1>\longrightarrow  |0>|1>,
 \nonumber \\ &&|1>|0>\longrightarrow  |1>|1>, \ \ |1>|1>\longrightarrow 
 |1>|0>, \label{cn1}
\end{eqnarray}
that is, it replaces the second bit by its opposite whenever the first 
bit is $|1>$, and leaves it untouched, when the first is $|0>$.
It is immediate that, classically, this operation clones the first 
component of a sequence of pairs of states (by imprinting it in the 
second component), when acting on a sequence of pairs where the second 
components are all 0's (blank sequence). Thus, for instance,
\begin{eqnarray}
&&|0>|0>, |0>|0>, |1>|0>, |0>|0>,  |1>|0>, |1>|0>, \ldots  \nonumber \\ 
 && \hspace*{15mm} \longrightarrow 
|0>|0>, |0>|0>, |1>|1>, |0>|0>,  |1>|1>, |1>|1>, \ldots \label{cn2}
\end{eqnarray}
Moreover, this operation is reversible: when confronted with a sequence of
identical pairs of states (such as for instance the second one in 
(\ref{cn2})),
the  C-not operation will turn out a sequence where the first 
components remain unchanged, while the second ones are all 0's.
It is rather immediate (see e.g. \cite{zn1})  that this simple 
operation fails
altogether as a copying or deleting tool when applied to pairs of 
{\it quantum} bits (qubits). In terms of the states $|0>$ and $|1>$ a 
qubit will have 
the general form $|s> = \alpha |0> + \beta  |1>$ ($\alpha$ and $\beta$ being 
complex numbers such that $|\alpha|^2 +|\beta|^2 =1$) and, the C-not operation 
acting, for instance, on the pair $|s>|0>$ yields an entangled state:
\begin{eqnarray} 
(\alpha |0> + \beta  |1>)|0> \longrightarrow \alpha |0>|0> + \beta  |1> |1>  
\label{cn3}
\end{eqnarray}
(the C-not operation is linear).
Put it in this way, it looks like there is no way out of the conclusion 
in the paper by  Pati and Braunstein \cite{pbn1} $-$to which the reader is 
addressed for conventions, notation and
additional references$-$ that it is impossible
to delete a copy of a quantum state, that comes in at least two copies,
$|s>|s>$. Our main idea here will be to use a {\it different} 
form of the  C-not 
operation together with an alternative notion of deletion, which 
depart from and extend in a way the definitions above. 

We shall now fix our strategy. We  give the term {\it deletion}  
a similar sense as in Ref. \cite{pbn1}, as has been just described and
 which, in the words of Zurek \cite{zn1}, may be termed a narrow
concept, but one that is being indeed most widely employed. This means, in
what follows we shall just (as above) delete a
single copy of some classical or quantum information piece that is kept in
some device in at least two copies, so that at least one copy of the 
information remains in the end.
But, as already advanced, 
our logical  deleting/cloning operations will be {\it different} 
from the C-not operation. They will be called, respectively,  {\it
random deletion} and {\it cloning from random}, the first being a
genuine R-choice. As it happens with the C-not operation, they also 
can operate both on classical and quantum systems. 
Being more specific, classical R-deletion will be defined as:
\begin{eqnarray}
\left\{ \begin{array}{l}
|0> \longrightarrow  R |0> = \left\{  \begin{array}{l} |0>, \ \ 
\mbox{with}\ \ p=1/2, \\
|1>, \ \ \mbox{with} \ \ q=1-p=1/2, \end{array} \right. \\
|1> \longrightarrow  R |1> = \left\{  \begin{array}{l} |0>,\ \ 
\mbox{with}\ \ p=1/2, \\
|1>, \ \ \mbox{with} \ \ q=1-p=1/2, \end{array} \right.
\end{array} \right.
\label{rd1}
\end{eqnarray}
where $p$ and $q$ denote probabilities and we have obviated the extra copies.
That is, any state of the classical system is replaced by a reference state,
which is not determined to be the $|0,0,0, \ldots, 0>$ state, but rather one
chosen each time at random, from a given set of pure states. 
Plainly, we would, for instance, delete the
information contained in a Shakespearian play not by replacing all the
characters with say $a$'s, but by replacing it with say\footnote{I apologize 
for the example} an equally long play written by a
monkey sitting in front of a typewriter (that is, an arbitrary state in a
thermal bath, rather than a chosen, standard, zero-state
\cite{zb2,zb1}).
Of course, it will not always be the same play
and the obvious question could be asked: how can we know {\it a priori},
on looking to a track or a whole disk, that it has undergone a deletion
process, e.g., that it is empty and does not contain any information, if 
spins are
 not all aligned as in the usual $|0,0,0, \ldots, 0>$ state? One of the 
possible answers: we know that the device is `empty' because it is labeled 
as such. For this we will just need an additional bit. To give a further 
visual picture of this last issue, let us mention that it freely corresponds 
to what we do at home when we decide that a video-tape or a CD are ready 
for re-use, after we have seen the
movie (or are just fed up with the music) that we had previously recorded 
on them. In short, the new bit just mentioned should mimic this practice.

It should be noted that there is presently a great deal of confusion about
the definition of `deletion' in the case of quantum information \cite{refe2}.
Classically, the convention seems to be that deletion is the
reversible implementation of the erasure map: $|\psi><\psi|\otimes
|\psi><\psi|\longrightarrow |\psi><\psi|\otimes|0><0|$ $-$where $|0>$ is a 
standard initial pure state$-$ with $|\psi>$ restricted to the classical 
symbols from some alphabet (usually $\{0,1\}$). Whether the map is 
reversible or not is determined by its action on states with different 
symbols on the two systems. Quantumly, it is obvious that this is not 
possible \cite{refe2}. A related (albeit different) definition of 
a (complete) randomization process, that has features in common with 
the one we have implemented in R-deletion, was introduced in 
\cite{BLS}, and 
has been used very recently in different situations \cite{MTW,BR,refe2}  
(although never in the context of the no-deletion principle). 
Ideally one would wish to delete without generating heat, but the amount
produced in the present case is strictly.
In particular, it has been proven in \cite{MTW} that the generation of 
$n$ bits of entropy is sufficient and necessary in order to randomize 
arbitrary tensor products of $n$ real amplitude qubits. 
Observe that this is just an upper bound, in our case, since the randomization
here amounts only to a unit fraction of the total for the tensor product 
state. While this might be a technical problem at some point, it does not 
prevent the definition and use of our R-deletion operator as a working, 
alternative erasure method and a substitute for ordinary deletion. (Often the 
physical implementation of related procedures is not without difficulties). 

One may observe that the purpose of deletion or erasure 
is to prepare the system for reuse, what requires a standard pure 
state, and that R-deletion (classically or quantumly) will defer this 
process to when the system is actually used again. And then, 
in the quantum case, the net effect of losing a copy of the state to 
the environment or the apparatus by the time the system is reused could
still occur. Neither of these are true. In fact, the use of a set of 
pure states for the family of 
standard states and the introduction of a new operator, that we shall call 
R-cloning $-$and is classically an extension of the ordinary
logical operation `inverse' to C-not$-$ will take care of the first issue.
As for the second, it turns out that the copy that is lost to the 
environment, by the time the system is reused, is going to be always an
informationless,
randomized remnant of the initial copy and {\it no} information will go to
the reusing apparatus either. The new pure state to be prepared for reuse is
got from the newly randomized state, which has lost any information on the 
initial one. Quantumly this corresponds to a diagonalization of
the randomized density matrix, obtained in the first step of the procedure. 

R-cloning has no essential difference with respect to the ordinary
logical operation `inverse' to C-not in Eq. (\ref{cn1}) \cite{wz1,d1}
(see also the very clear description by Zurek
\cite{zn1}). By presenting pairwise the state to be cloned together with the
random state (whatever it may be), R-cloning will replace the last
with the first. There is nothing essentially distinct with the procedure
in this case but, rigorously, it is a {\it different} operator, as we 
shall later explain. 

In complete analogy with (\ref{rd1}), quantum R-deletion starts 
by considering first a family of quantum states, labeled by a random 
parameter, $\sigma$. The parameter range can be finite or infinite, 
discrete or continuous.
Actually, a family with only two members (as the states of a pair of 
horizontally and vertically polarized photons)
 would suffice to get an information loss in
the way we are going to see. (For an example of a continuous, infinite family 
one may think, for instance, in the set of wave functions $\varphi_\sigma (x)$
given by Gaussian distributions N(0, $\sigma$), with $\sigma$ a 
positive real number that labels here the  family of quantum states
$|\varphi_\sigma >$). The quantum states in the family will be
chosen to be pure states. Using the same notation of Pati and Braunstein
\cite{pbn1}, let us consider a couple of identical qubits (as two photons of
arbitrary  polarization) in some quantum state $| \psi >$ together with an
ancilla in a state $|A>$, corresponding to the `ready' state of the 
deleting device \cite{pbn1}. The aim of the deleting device in the spirit of 
Ref. \cite{pbn1} was to replace one of the two copies of $| \psi >$ with 
some standard, fixed state of a qubit $|\Sigma >$. However,  in the spirit 
of our approach to deletion (as described above, in the classical 
case), the R-deletion operator yields, in the quantum situation:
\begin{eqnarray}
|\psi >|\psi > |A> \longrightarrow     |\psi >|\Sigma_{\sigma_1}>|A_\psi >,
\label{ppa1}
\end{eqnarray}
where now the standard state of a qubit ($|\Sigma >$ in \cite{pbn1}) is
replaced with one of the family of pure states,  chosen at random,
$|\Sigma_{\sigma_1}>$.

When we now consider the action of R-deletion on a pair of horizontally
and vertically polarized photons, respectively, we obtain
\begin{eqnarray}
|H >|H > |A> &\longrightarrow &    |H >|\Sigma_{\sigma_2}>|A_H >,
\label{hha1} \\
|V >|V > |A> &\longrightarrow &    |V >|\Sigma_{\sigma_3}>|A_V >,
\label{vva1}
\end{eqnarray}
and odds are against having the same standard state for the deleted
copy, namely $|\Sigma_{\sigma}>$ (therefore the three different labels).
By applying now
R-deletion to an arbitrary input qubit $|\psi > =\alpha |H > + \beta |V>$
(with $|H>$ and $|V>$ forming a basis, and $\alpha$ and $\beta$ being 
complex numbers such that $|\alpha|^2 +|\beta|^2 =1$), we obtain
\begin{eqnarray}
|\psi >|\psi > |A> \longrightarrow    \alpha^2|H >|\Sigma_{\sigma_2}>|A_H > +
\beta^2|V >|\Sigma_{\sigma_3}>|A_V > + \sqrt{2} \ \alpha \beta |\Phi>,
\end{eqnarray}
where $|\Phi>$ is the state obtained by R-deletion of the entangled state
$(1/\sqrt{2})(|H>|V> + |V>|H>)|A>$. But, whatever  this state be in our actual
realization of R-deletion, we do {\it not} recover now, from the linearity
of Quantum Mechanics,  the result in \cite{pbn1}
that $|A_\psi> = \alpha |A_H> +\beta |A_V>$. In fact, the strong 
implications that the linearity of QM has for the ordinary deletion process 
are {\it lost} when using the R-deletion transformation (the most obvious 
reason being the lack of a unique, standard $|\Sigma>$ state to play with). 
Note that, in order to completely define the operation, we must still 
assign values to the rest of the basis of the tensor space, namely:
$|H>|V>|A>$ and  $|V>|H>|A>$ $-$what can be done freely$-$ to close then the 
definition by extending {\it linearly} the operator to the whole of the 
tensor space of our quantum system \cite{kato}. This is in common with 
the ordinary definition of 
deletion, the only difference being thus the multiple choice of pure states 
to be assigned to the last component of the symmetrical elements of the basis 
for the tensor product. 
The generalization of this example to higher dimensional Hilbert spaces is
straightforward.

In few words, the definition of R-deletion, although  linear in essence, 
induces a weird behaviour when combined with the linear transformations of 
QM.  As it involves randomly several `standard' states, it
prevents the linearity of the relation $|\psi > =\alpha 
|H > + \beta |V>$ from being simply transfered to the ancilla state $|A_\psi>$,
what would eventually lead to a preservation of the whole information content
and prevent its deletion \cite{pbn1}. In fact, R-deletion acting on
$|\psi >|\psi > |A>$ yields (\ref{ppa1}) which, on the other hand, if
$|A_\psi>$ were to be given by $|A_\psi> = \alpha |A_H> + \beta |A_V>$, by
taking into account (\ref{hha1}) and (\ref{vva1}), would yield the identity:
\begin{eqnarray}
&& \alpha^2|H >|\Sigma_{\sigma_2}>|A_H > +
\beta^2|V >|\Sigma_{\sigma_3}>|A_V > + \sqrt{2} \alpha \beta |\Phi> 
\nonumber \\
 &&  \hspace{10mm}  =
\alpha^2|H >|\Sigma_{\sigma_1}>|A_H > +
\beta^2|V >|\Sigma_{\sigma_1}>|A_V > \\ && \hspace{16mm} + \alpha \beta
\left( |H >|\Sigma_{\sigma_1}>|A_V > +
|V >|\Sigma_{\sigma_1}>|A_H > \right), \nonumber
\end{eqnarray}
which is impossible to fulfill given the nature of R-deletion (in particular,
the randomness of the $\sigma_i$, $i=1,2,3$). Given that the states
$|\Sigma_{\sigma_i}>$, $i=1,2,3$, are unrelated, it is now
impossible to recover information about the state $|\psi>$ from
that of the ancilla remnant $|A_\psi>$. In conclusion, we have actually 
managed to delete one copy of the information, in the end.

R-deletion does not seem to have much new to say on results about cloning.
Notwithstanding that, if one considers cloning as (in some sense) an 
inverse procedure to deleting,
we are here in the presence of two sensibly different deleting operations to
which, in the classical case, the same cloning operation appears to be a 
common inverse.
This is in fact not completely true. Strictly speaking, the cloning
operation which is the inverse of R-deletion strictly {\it contains} 
the cloning operation that is the
inverse of C-not. This is so, since the last
cloning will always act on the standard blank state $|0,0,0, \ldots, 0>$ or
$|\Sigma>$ state only (as the second state, to be cloned). Only provided its
definition were extended, allowing it to act on
{\it any} state as second in the pair,  would it coincide  with the
cloning which is the (classical) inverse of R-deletion.

To summarize, the new logical operation, R-deletion, that has been here
introduced, is able to avoid the linear
transference of the information kept in any arbitrary quantum state to the
ancilla, as it inevitably happened with the
ordinary deletion procedure \cite{pbn1}, what prevented even the least 
amount of deletion  of quantum information in that case. Our result provides
support to a recent suggestion \cite{refe2} that the 
non-deletion principle should preferably be called the erasure-only 
principle. 

\medskip

\noindent{\bf Acknowledgments}  
 
This investigation has been supported by DGICYT (Spain), project  
PB96-0925, by CIRIT (Ge\-ne\-ra\-li\-tat de Catalunya),
grant 1999SGR-00257, and by a USA-Spain Cooperation Agreement.
 
\vspace{1mm} 



\begin{thebibliography}{99}

\bibitem{pbn1} A.K. Pati and S.L. Braunstein, {\it Impossibility of
deleting an unknown quantum state}, Nature {\bf 404}, 164-165 (2000).

\bibitem{bb1} C.H. Bennett and G. Brassard,   {\it The dawn of a new era for 
quantum cryptography: the experimental prototype is working!}, SIGACT News
 {\bf 20}, 78-82 (1989).
 

\bibitem{cb1} C.H. Bennett et al., {\it Teleporting an unknown quantum 
state via dual classical and Einstein-Podolsky-Rosen channels},  
Phys. Rev. Lett. {\bf 70}, 1895-1899 (1993).  

\bibitem{db1} D. Bouwmeester et al., {\it  Experimental quantum teleportation},
 Nature {\bf 390}, 575-579 (1997).  

\bibitem{z1} A. Zeilinger, {\it Fundamentals of quantum information}, 
Phys. World (March), 35-40 (1998).

\bibitem{wz1} W.K. Wooters and W.H. Zurek, {\it A single quantum cannot
be cloned}, Nature {\bf 299}, 802-803 (1982).

\bibitem{d1} D. Dieks, {\it Communication by EPR devices}, Phys. Lett.
{\bf A92}, 271-272 (1982).

\bibitem{bh1} V. Bu$\check {z}$ek and M.H. Hillery,  {\it Quantum copying: 
beyond the no-cloning theorem},   Phys. Rev. A {\bf 54}, 1844-1852 (1996).

\bibitem{bh11} V. Bu$\check{z}$ek, S.L. Braunstein, M.H. Hillery, and D.
Bru{\ss}, {\it Quantum copying: a network},   Phy. Rev. A {\bf 56}, 
3446-3452 (1997).  

\bibitem{gm1} N. Gisin and S. Massar, {\it Optimal quantum cloning machines},
 Phys. Rev. Lett. {\bf 79}, 2153-2156 (1997).  

\bibitem{dg2} L.M. Duan and G.C.  Guo, {\it Probabilistic cloning and 
identification of linearly independent states}, Phys. Rev. Lett. {\bf 80}, 
4999-5002 (1998).  

\bibitem{y1} H.P. Yuen, {\it Amplification of quantum states and noiseless 
photon amplifiers},  Phys. Lett. A {\bf 113}, 405-407 (1986).     

\bibitem{p1} A.K. Pati, {\it Quantum superposition of multiple clones and the
novel cloning machine},  Phys. Rev. Lett. {\bf 83}, 2849-2852 (1999). 

\bibitem{sz1} L. Szilard, Z. Phys. {\bf 53}, 840-856 (1929). 

\bibitem{la1} R. Landauer, IBM J. Res. Dev. {\bf 3}, 183-191 (1961).

\bibitem{zn1} W.H. Zurek, {\it Schr\"odinger's sheep}, Nature {\bf 404},
130-131 (2000).
  
\bibitem{zb2} E. Elizalde, {\it Ten physical  
applications of spectral zeta functions} (Springer, Berlin, 1995).

\bibitem{zb1} E. Elizalde, S.D. Odintsov, A.  
Romeo, A.A. Bytsenko and S. Zerbini, {\it Zeta regularization  
techniques with applications} (World Sci., Singapore, 1994).  
 
\bibitem{refe2} Thanks are given to an anonimous referee for this crucial
observation.   

\bibitem{BLS} S. Braunstein, H.-K. Lo, and T. Spiller, {\it Forgetting
qubits is hot to do} (1999), unpublished.

\bibitem{MTW} M. Mosca, A. Tapp, and R. de Wolf, {\it  Private quantum 
channels and the cost of randomizing quantum information}, quant-ph/0003101.
\bibitem{BR} P.O. Boykin and V. Roychowdhury, {\it Optimal encryption of 
quantum bits}, quant-ph/0003059.

\bibitem{kato} T. Kato, {\em Perturbation theory for linear operators} 
(Springer, Berlin, 1980).    

 
\end{thebibliography}
\end{document}